\documentclass[preprint,12pt]{elsarticle}
\usepackage{amsthm}
\usepackage{enumerate}
\usepackage{color}
\usepackage{amsfonts}
\usepackage{tabularx}
\usepackage{mathrsfs}
\usepackage{mathdots}
\usepackage{verbatim}
\usepackage{graphicx}
\usepackage{subfigure}
\usepackage{amsmath}
\usepackage{mathabx}
\usepackage{multirow}
\usepackage{float}
\usepackage{amssymb}%

\newcounter{algorithm}
\newenvironment{algorithm}{\refstepcounter{algorithm}\vspace{1ex}
  {\sc Algorithm \thealgorithm.}\hspace{0.3em}\parindent=0pt}{\vspace{1ex}}
\newcounter{remark}
\newenvironment{remark}{\refstepcounter{remark}\vspace{1ex}
{\sc Remark \theremark.}\hspace{0.3em}\parindent=0pt}{\vspace{1ex}}

\theoremstyle{plain}

\def\diag{\text{diag}}

\usepackage[%
  breaklinks=true,%
  colorlinks=true,%
  linkcolor=blue,anchorcolor=blue,%
  citecolor=blue,filecolor=blue,%
  menucolor=blue,pagecolor=blue,%
  urlcolor=blue]{hyperref}

\journal{J. Comput. Appl. Math.}

\begin{document}

\begin{frontmatter}



\title{An efficient hybrid tridiagonal divide-and-conquer algorithm on distributed memory architectures}

\author[work1]{Shengguo Li\corref{cor}}
\cortext[cor]{Corresponding author} \ead{nudtlsg@nudt.edu.cn}
\author[work2]{Fran\c{c}ois-Henry Rouet}
\author[work1,work3]{Jie Liu}
\author[work1,work3]{Chun Huang}
\author[work4]{Xingyu Gao}
\author[work5]{Xuebin Chi}


\address[work1]{College of Computer, National University of Defense Technology (NUDT), Changsha 410073, China}

\address[work2]{Lawrence Berkeley National Laboratory, Berkeley, CA 94720, USA}

\address[work3]{State Key Laboratory of High Performance Computing, NUDT, China}

\address[work4]{Institute of Applied Physics and Computational Mathematics, Beijing 100094, China}

\address[work5]{Computer Network Information Center, Chinese Academy of Science, Beijing 100190, China}

\begin{abstract}
  In this paper, an efficient divide-and-conquer (DC) algorithm is proposed for the symmetric tridiagonal matrices
  based on ScaLAPACK and the hierarchically semiseparable (HSS) matrices.
  HSS is an important type of rank-structured matrices.
  Most time of the DC algorithm is cost by computing the eigenvectors via
  the matrix-matrix multiplications (MMM).
  In our parallel hybrid DC (PHDC) algorithm, MMM is accelerated by using the HSS matrix
  techniques when the intermediate matrix is large.
  All the HSS algorithms are done via the package \texttt{STRUMPACK}.
  PHDC has been tested by using many different matrices.
  Compared with the DC implementation in MKL,
  PHDC can be faster for some matrices with few deflations when using
  hundreds of processes.
  However, the gains decrease as the number of processes increases.
  The comparisons of PHDC with ELPA (the Eigenvalue soLvers for Petascale Applications library) are similar.
  PHDC is usually slower than MKL and ELPA when using 300 or more processes
  on Tianhe-2 supercomputer.
\end{abstract}

\begin{keyword}
ScaLAPACK \sep Divide-and-conquer \sep HSS matrix \sep Distributed parallel algorithm
\MSC 65F15, 68W10
\end{keyword}
\end{frontmatter}

\section{Introduction}
\label{sec:intro}

The symmetric tridiagonal eigenvalue problems are usually solved by the divide and
conquer (DC) algorithm both on shared memory multicore platforms and parallel distributed memory machines.
The DC algorithm is fast and stable, and well-studied in numerous references
\cite{Cuppen81,BNS-Rankone,Ipsen-TDC,Gu-eigenvalue,Gates-Arbenz,Tisseur-DC}.
It is now the default method in LAPACK~\cite{anderson1999lapack} and
ScaLAPACK~\cite{Scalapack} when the eigenvectors
of a symmetric tridiagonal matrix are required.

Recently, the authors~\cite{LSG-NLAA} used the hierarchically semiseparable (HSS) matrices~\cite{ChandrasekaranGu04}
to accelerate the tridiagonal DC in LAPACK, and obtained about 6x
speedups in comparison with that in LAPACK for some large matrices on a shared memory multicore platform.
The bidiagonal and banded DC algorithms for the SVD problem are accelerated similarly~\cite{Shengguo-SIMAX2,Liao-camwa}.
The main point is that some intermediate eigenvector matrices are rank-structured matrices
\cite{ChandrasekaranGu04,Hackbusch1999}.
The HSS matrices are used to approximate them and then use fast HSS algorithms to update the eigenvectors.
HSS is an important type of rank-structured matrices, and others include $\mathcal{H}$-matrix \cite{Hackbusch1999},
$\mathcal{H}^{2}$-matrix \cite{Hackbusch-Sauter2000}, quasiseparable~\cite{Eidelman-Gohberg1999} and
 sequentially semiseparable (SSS)~\cite{CG-SSS-report,Gu-DSSS}.
In this paper, we extend the techniques used in~\cite{Shengguo-SIMAX2,LSG-NLAA} to the distributed memory environment,
try to accelerate the tridiagonal DC algorithm in ScaLAPACK~\cite{Scalapack}.
To integrate HSS algorithms into ScaLAPACK routines, an efficient distributed HSS construction routine
and an HSS matrix multiplication routine are required.
In our experiments, we use the routines in STRUMPACK (STRUctured Matrices PACKage) package~\cite{Strumpack},
which is designed for computations with both \emph{sparse}
and \emph{dense} structured matrices.
The current STRUMPACK has two main components: \emph{dense matrix computation package}
and \emph{sparse direct solver and preconditioner}.
In this work we only use its dense matrix operation part\footnote{The current version is STRUMPACK-Dense-1.1.1,
which is available at http://portal.\allowbreak nersc.gov/project/sparse/strumpack/}.
It is written in C++ using OpenMP and MPI parallism,
uses HSS matrices, and it implements a parallel HSS construction algorithm
with randomized sampling~\cite{Martinsson-randhss10,Martinsson-Rev10}.
Note that some routines are available for sequential HSS algorithms~\cite{Xia-random,SSS-QR} or
parallel HSS algorithms on shared memory platforms
such as HSSPACK~\cite{Liao-camwa}\footnote{Some Fortran and Matlab codes are available at Jianlin Xia's homepage,
http://\allowbreak www.\allowbreak math.purdue.edu/\~{}xiaj/, and HSSPACK is available at GitHub.}.
But STRUMPACK is the only available one for the distributed parallel HSS algorithms.
More details about it and HSS matrices will be introduced in section~\ref{sec:hss}.

The ScaLAPACK routine implements the rank-one update of Cuppen's DC algorithm~\cite{Cuppen81}.
We briefly introduce the main processes.
Assume that $T$ is a symmetric tridiagonal matrix,
\begin{equation}
  \label{eq:Tmat}
  T=
  \begin{pmatrix}
    a_1 & b_1 & & \\
    b_1 & \ddots & \ddots & \\
    & \ddots & a_{N-1} & b_{N-1} \\
    & & b_{N-1} & a_N
  \end{pmatrix}.
\end{equation}
Cuppen introduced the decomposition
\begin{equation}
  \label{eq:T2}
  T=
  \begin{pmatrix}
    T_1 & \\ & T_2
  \end{pmatrix}+b_k vv^T,
\end{equation}
where $T_1\in R^{k\times k}$ and $v=[0,\ldots,0,1,1,0,\ldots,0]^T$ with
ones at the $k$-th and $(k+1)$-th entries.
Let $T_1=Q_1 D_1Q_1^T$ be $T_2=Q_2 D_2 Q_2^T$ be eigen decompositions, and then~\eqref{eq:Tmat} can be written as
\begin{equation}
  \label{eq:T3}
  T=
  Q \left(
  D + b_k zz^T \right)
Q^T,
\end{equation}
where $Q=\diag(Q_1, Q_2)$, $D=\diag(D_1, D_2)$ and $z=
Q^Tv =
\begin{pmatrix}
  \text{last column of } Q_1^T \\
  \text{first column of } Q_2^T
\end{pmatrix}.
$
The problem is reduced to computing the spectral decomposition of the diagonal plus rank-one
\begin{equation}
\label{eq:rankone}
D+b_k uu^T=\widehat{Q}\Lambda \widehat{Q}^T.
\end{equation}
By Theorem 2.1 in~\cite{Cuppen81}, the eigenvalues $\lambda_i$ of matrix $D+b_kzz^T$ are the root of the secular equation
\[
f(\lambda)=1+b_k \frac{z_k^2}{d_k-\lambda} = 0,
\]
where $z_k$ and $d_k$ are the $k$th component of $z$ and the \emph{k}th diagonal entry of $D$, respectively,
and its corresponding eigenvector is given by $\hat{q}_i=(D-\lambda_i)^{-1}z.$
The eigenvectors simply computed this way may loss orthogonality.
To ensure orthogonality, Sorensen and Tang~\cite{Sorensen-sina91} proposed to use extended precision.
While, the implementation in ScaLAPACK uses the L\"{o}wner theorem approach,
instead of the extended precision approach~\cite{Sorensen-sina91}.
The extra precision approach was used by Gates and Arbenz~\cite{Gates-Arbenz} in their implementation.

\begin{remark}
The extra precision approach is ``embarrassingly" parallel with each eigenvalue and eigenvector
computed without communication, but it is not portable in some platform.
The L\"{o}wner approach requires information about all the eigenvalues, requiring a broadcast.
However, the length of communication message is $O(n)$ which is trivial compared with the $O(n^2)$ communication
of eigenvectors.
\end{remark}

The excellent performance of the DC algorithm is partially due to \emph{deflation}~\cite{BNS-Rankone,Cuppen81}, which
happens in two cases.
If the entry $z_i$ of $z$ are negligible or zero, the corresponding $(\lambda_i, \hat{q}_i)$ is already an
eigenpair of $T$.
Similarly, if two eigenvalues in $D$ are identical then one entry of $z$ can be transformed to zero by
applying a sequence of plane rotations.
All the deflated eigenvalues would be permuted to back of $D$ by a permutation matrix, so do the
corresponding eigenvectors.
Then~\eqref{eq:T3} reduces to, after deflation,
\begin{equation}
\label{eq:T3-def}
T=Q(GP)\begin{pmatrix}\bar{D}+b_k\bar{z}\bar{z}^T & \\ & \bar{D}_d \end{pmatrix}
(GP)^T Q^T,
\end{equation}
where $G$ is the product of all rotations, and $P$ is a permutation matrix and
$\bar{D}_d$ are the deflated eigenvalues.

According to~\eqref{eq:rankone}, the eigenvectors of $T$ are computed as
\begin{equation}
\label{eq:evect}
U = Q(GP)\begin{pmatrix}\widehat{Q} & \\ & I_d  \end{pmatrix} =
\left[ \begin{pmatrix} Q_1 & \\ & Q_2 \end{pmatrix} GP \right] \begin{pmatrix} \widehat{Q} & \\ & I_d \end{pmatrix}.
\end{equation}
To improve efficiency, Gu~\cite{Gu-thesis} suggested a permutation strategy for reorganizing
the data structure of the orthogonal matrices, which has been used in ScaLAPACK.
The matrix in square brackets is permuted as
$\begin{pmatrix}{Q}_{11} & {Q}_{12} & 0 & {Q}_{14} \\
0 & {Q}_{22} & {Q}_{23} & {Q}_{24} \end{pmatrix}$,
where the first and third block columns contain the eigenvectors that have not been affected
by deflation, the fourth block column contains the deflated eigenvectors, and the second block
column contains the remaining columns.
Then, the computation of $U$ can be done by two parallel matrix-matrix products
(calling PBLAS \texttt{PDGEMM}) involving parts of $\widehat{Q}$ and
the matrices $\begin{pmatrix} Q_{11} & Q_{12} \end{pmatrix}$,
$\begin{pmatrix} Q_{22} & Q_{23} \end{pmatrix}$.
Another factor that contributes to the excellent performance of DC is that
most operations can take advantage of highly optimized matrix-matrix products.

When there are few deflations, the size of matrix $\widehat{Q}$ in~\eqref{eq:evect} will be large, and
most time of DC would be cost by the matrix-matrix multiplication in~\eqref{eq:evect}, which
is confirmed by the results of Example 2 in section~\ref{sec:hsseig}.
Furthermore, it is well-known that matrix $\widehat{Q}$ defined as in~\eqref{eq:rankone} is a Cauchy-like
matrix with off-diagonally low rank property, see~\cite{Gu-eigenvalue,LSG-NLAA}.
Therefore, we simply use an HSS matrix to approximate
$\widehat{Q}$ and use HSS matrix-matrix multiplication routine in STRUMPACK to
compute the eigenvector matrix $U$ in~\eqref{eq:evect}.
Since HSS matrix multiplications require much fewer floating point operations than
the plain matrix-matrix multiplication, \texttt{PDGEMM}, this approach
makes the DC algorithm in ScaLAPACK much faster.
More details are included section~\ref{sec:hsseig} and numerical results are shown
in section~\ref{sec:num}.

\section{HSS matrices and STRUMPACK}
\label{sec:hss}

The HSS matrix is an important type of rank-structured matrices.
A matrix is called \emph{rank-structured} if the ranks of all off-diagonal
blocks are relatively small compared to the size of matrix.
Other rank-structured matrices include  $\mathcal{H}$-matrix \cite{Hackbusch1999,Hackbusch2000},
$\mathcal{H}^{2}$-matrix \cite{Hackbusch-Sauter2000,Hackbusch-Borm2002},
quasiseparable matrices \cite{Eidelman-Gohberg1999, Vandebril-book1}, and sequentially
semiseparable (SSS) \cite{Chandrasekaran03,ChandrasekaranGu05} matrices.
We mostly follow the notation used in~\cite{Martinsson-randhss10} and~\cite{Xia-Fast09,LSG-NLAA}
to introduce HSS.

\begin{figure}
\centering
\subfigure[Matrix $A$]{
\includegraphics[width=2.0in]{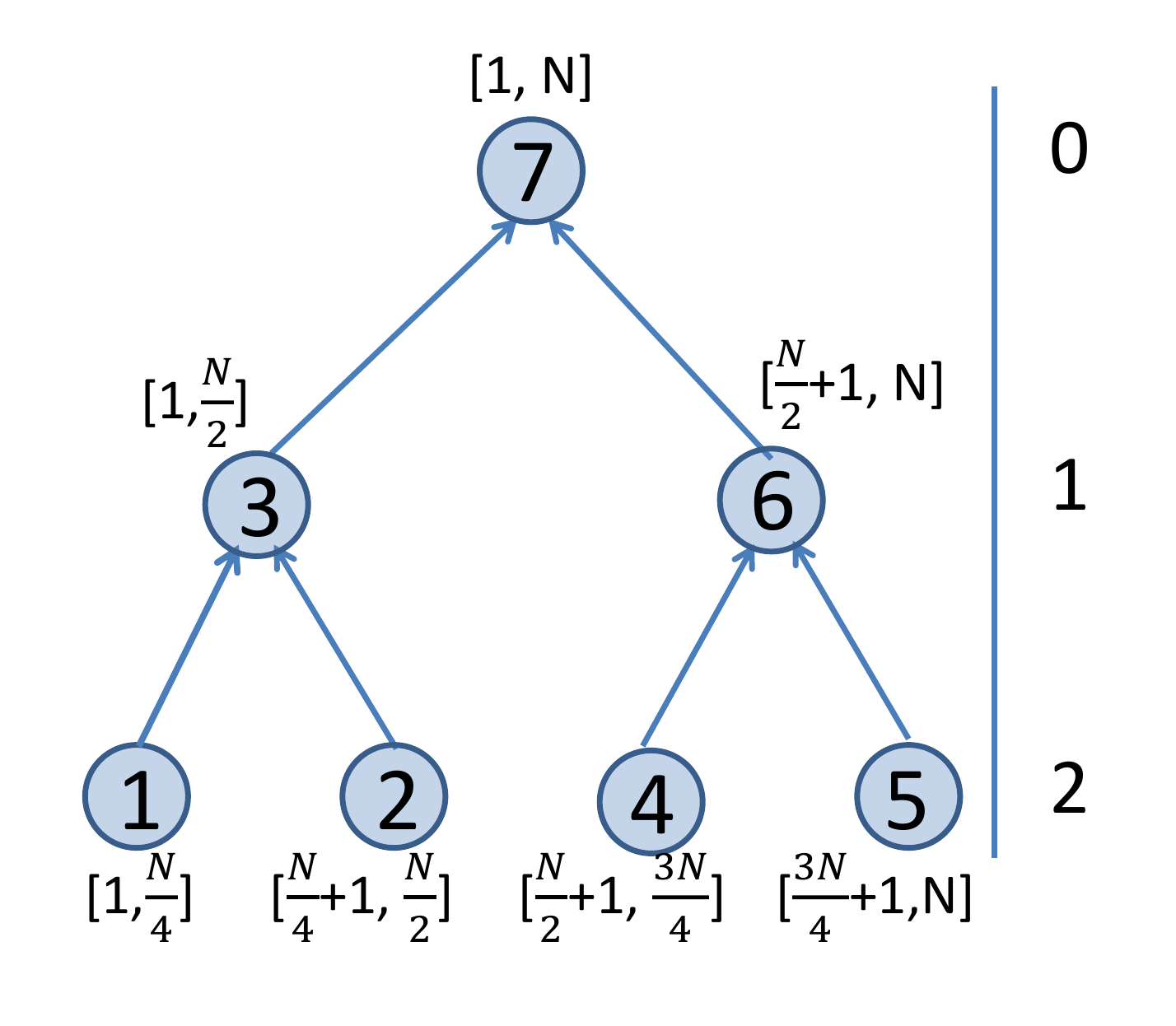}
\label{fig:htree}}
\quad
\subfigure[HSS tree]{
\includegraphics[width=2.1in]{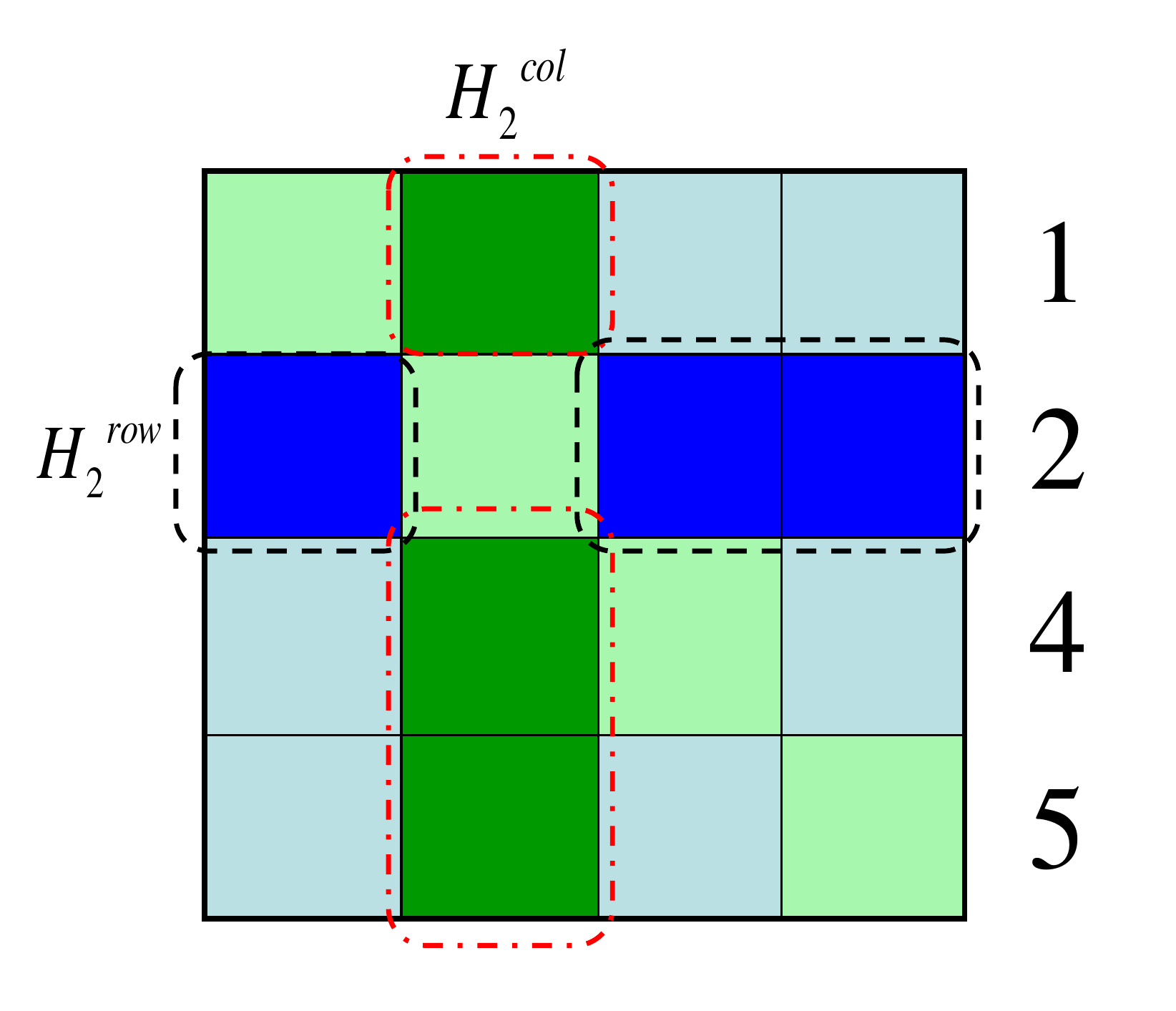}
\label{fig:Hssblock}}
\caption{A three level postordering tree $\mathcal{T}$ and the HSS blocks of Node 2}
\label{fig:gnaHss}
\end{figure}

Both HSS representations and algorithms rely on a \emph{tree} $\mathcal{T}$.
For simplicity, we assume it is a \emph{binary tree}, name it \emph{HSS tree},
and the generalization is straightforward.
Assume that $\mathcal{I}=\{1,2,\ldots, N\}$  and $N$ is the dimension of matrix $A$.
Each node $i$ of $\mathcal{T}$ is associated with a contiguous subset of $\mathcal{I}$, $t_i$, satisfying the
following conditions:
\begin{itemize}
\item $t_{i_1}\cup t_{i_2}=t_i$ and $t_{i_1}\cap t_{i_2}=\emptyset$, for a parent node $i$ with
  left child $i_1$ and right child $i_2$;

  \item $\cup_{i \in LN} t_i = \mathcal{I}$, where $LN$ denotes the set of all leaf nodes;

\item $t_{\operatorname*{root}(\mathcal{T})}=\mathcal{I}$,
  $\operatorname*{root}(\mathcal{T})$ denotes the root of $\mathcal{T}$.
\end{itemize}
We assume $\mathcal{T}$ is \emph{postordered}.
That means the ordering of a nonleaf node $i$ satisfies $i_1 < i_2 < i$,
where $i_1$ is its left child and $i_2$ is its right child.
Figure~\ref{fig:htree} shows an HSS tree with three levels
and $t_i$ associated with each node $i$.

A block row or column excluding the diagonal block is called an \emph{HSS block row or column},
denoted by
\[
H_i^{row} = A_{t_i \times (\mathcal{I}\backslash t_i)}, \quad H_i^{col} = A_{(\mathcal{I}\backslash t_i)\times t_i},
\]
associated with node $i$. We simply call them \emph{HSS blocks}.
For an HSS matrix, HSS blocks are assumed to be numerically low-rank.
Figure~\ref{fig:Hssblock} shows the HSS blocks corresponding to node 2.
We name the maximum (numerical) rank of all the HSS blocks by \emph{HSS rank}.

For each node $i$ in $\mathcal{T}$, it associates with four \emph{generators} $\widehat{D}_i$, $\widehat{U}_i$, $\widehat{V}_i$ and $B_i$,
which are matrices, such that
\begin{equation}
  \label{eq:new-hss}
  \begin{split}
  \widehat{D}_i & =A|_{t_i\times t_i}=
  \begin{bmatrix} \widehat{D}_{i_1} & \widehat{U}_{i_1}B_{i_1}\widehat{V}_{i_2}^T \\
    \widehat{U}_{i_2} B_{i_2} \widehat{V}_{i_1}^T & \widehat{D}_{i_2} \end{bmatrix}, \\
  \widehat{U}_i & =
  \begin{bmatrix}
    \widehat{U}_{i_1} & \\ & \widehat{U}_{i_2}
  \end{bmatrix}
  U_i, \quad
  \widehat{V}_i =
  \begin{bmatrix}
    \widehat{V}_{i_1} & \\ & \widehat{V}_{i_2}
  \end{bmatrix} V_i.
  \end{split}
\end{equation}
For a leaf node $i$, $\widehat{D}_i=D_i$, $\widehat{U}_i = U_i$,
$\widehat{V}_i=V_i$.
A $4\times 4$ (block) HSS matrix $A$ can be written as
\begin{equation}
\label{eq:posthsslevel}
A=\begin{bmatrix} \begin{bmatrix} D_1 & {U}_1B_1{V}_2^T \\  {U}_2B_2{V}_1^T & D_2 \end{bmatrix}   & \widehat{U}_3B_3\widehat{V}_6^T \\
\widehat{U}_6B_6\widehat{V}_3^T &
\begin{bmatrix} D_4 & {U}_4B_4{V}_5^T \\  {U}_5B_5{V}_4^T & D_5 \end{bmatrix}  \end{bmatrix},
\end{equation}
and Figure~\ref{fig:htree} shows its corresponding postordering HSS tree.
The HSS representation~\eqref{eq:new-hss} is equivalent to the representations in~\cite{Strumpack,Hss-ulv,ChandrasekaranGu04,Xia-HSS-Chol}.

To take advantage of the fast HSS algorithms, we need to construct
an HSS matrix first.
There exist many HSS construction algorithms
such as using SVD~\cite{Hss-ulv}, RRQR(rank revealing QR)~\cite{Xia-Fast09}, and so on.
Most of these algorithms cost in $O(N^2r)$ flops, where $N$ is the dimension of matrix and
$r$ is its HSS rank.
In~\cite{Martinsson-randhss10}, a randomized HSS construction algorithm (\texttt{RandHSS})  is proposed,
which combines random sampling with \emph{interpolative decomposition} (ID),
see~\cite{Cheng-Random,Martinsson-Rev10,Martinsson-PNAS07}.
The cost of \texttt{RandHSS} can be reduced to $O(Nr)$ flops if
there exists a fast matrix-vector multiplication algorithm in order of $O(N)$ flops.
STRUMPACK also uses this algorithm to construct HSS matrices.
For completeness, \texttt{RandHSS} is restated in Algorithm~\ref{alg:randhss}.

\begin{algorithm}
\label{alg:randhss}
(Randomized HSS construction algorithm)
Given a matrix $A$ and integers $r$ and $p$,
generate two $N\times (r+p)$ Gaussian random matrices
$\Omega^{(1)}$ and $\Omega^{(2)}$, and then compute
matrices $Y=A\Omega^{(1)}$ and $Z=A^T \Omega^{(2)}$.

\begin{description}

\item \texttt{do} $\ell=L,\cdots,1$
  \item[\quad] \texttt{for} node $i$ at level $\ell$
    \item[\qquad] \texttt{if} $i$ is a leaf node,
      \begin{enumerate}
        \item $D_i=A_{t_i,t_i}$;

        \item compute $\Phi_i=Y_i-D_i\Omega_i^{(1)}$, $\Theta_i=Z_i-D_i^T \Omega_i^{(2)}$;

        \item compute the ID of  $\Phi_i \approx U_i \Phi_i|_{\tilde{I}_i}$, $\Theta_i \approx V_i \Theta_i|_{\tilde{J}_i} $;

        \item compute $\widehat{Y}_i = V_i^T \Omega_i^{(1)}$, $\widehat{Z}_i = U_i^T \Omega_i^{(2)}$;
       \end{enumerate}

    \item[\qquad] \texttt{else}
      \begin{enumerate}

        \item store generators $B_{i_1} = A(\tilde{I}_{i_1},\tilde{J}_{i_2})$, $B_{i_2} = A(\tilde{I}_{i_2},\tilde{J}_{i_1})$;

        \item compute $\Phi_i=\begin{bmatrix} \Phi_{i_1}|_{\tilde{I}_{i_1}}-B_{i_1}\widehat{Y}_{i_2} \\ \Phi_{i_2}|_{\tilde{I}_{i_2}}-B_{i_2}\widehat{Y}_{i_1} \end{bmatrix}$, \quad
          $\Theta_i=\begin{bmatrix} \Theta_{i_1}|_{\tilde{J}_{i_1}}-B_{i_2}^T\widehat{Z}_{i_2} \\ \Theta_{i_2}|_{\tilde{J}_{i_2}}-B_{i_1}^T\widehat{Z}_{i_1} \end{bmatrix}$;

        \item compute the ID of  $\Phi_i \approx U_i \Phi_i|_{\tilde{I}_i}$, \quad $\Theta_i \approx V_i \Theta_i|_{\tilde{J}_i}$;

        \item Compute $\widehat{Y}_i = V_i^T \begin{bmatrix} \widehat{Y}_{i_1} \\ \widehat{Y}_{i_2} \end{bmatrix}$,  \quad
                      $\widehat{Z}_i = U_i^T \begin{bmatrix} \widehat{Z}_{i_1} \\ \widehat{Z}_{i_2} \end{bmatrix}$;
       \end{enumerate}

    \item[\qquad] \texttt{end if}

   \item[\quad] \texttt{end for}

\item \texttt{end do}

\end{description}
For the root node $i$, store $B_{i_1} = A(\tilde{I}_{i_1},\tilde{J}_{i_2})$, $B_{i_2} = A(\tilde{I}_{i_2},\tilde{J}_{i_1})$.

\end{algorithm}

The parameter $r$ in Algorithm~\ref{alg:randhss} is an estimate of the HSS rank of $A$, which
would be chosen adaptively in STRUMPACK~\cite{Strumpack}, and
$p$ is the oversampling parameter, usually equals to $10$ or $20$, see~\cite{Martinsson-Rev10}.
After matrix $A$ is represented in its HSS form, there exist fast algorithms for
multiplying it with a vector in $O(Nr)$ flops~\cite{Lyons-thesis,Chandrasekaran03}.
Therefore, multiplying an HSS matrix with another $N\times N$ matrix only costs in
$O(N^2r)$ flops.
Note that the general matrix-matrix multiplication algorithm \texttt{PDGEMM}
costs in $O(N^3)$ flops.

\subsection{STRUMPACK}
\label{sec:strumpack}

STRUMPACK is an abbreviation of \emph{STRUctured Matrices PACKage},
designed for computations with sparse and dense structured matrices.
It is based on some parallel HSS algorithms using randomization~\cite{Strumpack,Strumpack-sparse}.

Comparing with ScaLAPACK, STRUMPACK requires more \emph{memory}, since besides
the original matrix it also stores the random vectors and the samples, and
the generators of HSS matrix.
The memory overhead increases as the HSS rank increases.
Therefore, STRUMPACK is suitable for matrices with low off-diagonal ranks.
For the our eigenvalue problems, we are fortunate that the HSS rank of
intermediate eigenvector matrices appeared in the DC algorithm is not large,
usually less than 100.
Through the experiments in section~\ref{sec:num},
we let the compression threshold of constructing HSS be $1.0e$-$14$,
to keep the orthogonality of computed eigenvectors.
One advantage of STRUMPACK is that it requires much fewer operations than the
classical algorithms by exploiting the off-diagonally low rank property.

Another property of STRUMPACAK, the same as other packages that
explores low-rank structures ($\mathcal{H}$-matrices~\cite{Hlib} and Block Low-Rank representations),
is irregular computational patterns, dealing with irregular and imbalanced
taskflows and manipulating a collection of small matrices instead of
one large one.
HSS algorithms requires a lower asymptotic complexity, but the
flop rate with HSS is often lower than with traditional matrix-matrix
multiplications, BLAS3 kernels.
We expect HSS algorithms to have good performances for problems with large size.
Therefore, we only use HSS algorithms when the problem size is large enough,
just as in~\cite{Shengguo-SIMAX2,LSG-NLAA}.

The HSS construction algorithm in STRUMPACK uses randomized sampling algorithm
combined with Interpolative Decomposition (ID)~\cite{Martinsson-PNAS07,Cheng-Random},
first proposed in \cite{Martinsson-randhss10}.
For this randomized algorithm, the HSS rank needs to be estimated in advance,
which is difficult to be estimated accurately.
An adaptive sampling mechanism is proposed in STRUMPACK, and
its basic idea~\cite{Strumpack} is `to start with a low number of random vectors $d$, and whenever the
rank found during Interpolative Decomposition is too large, $d$ is increased'.
To control the communication cost, we use a little more sample vectors ($p=100$) and
let \texttt{inc\_rand\_HSS} relatively large which is a parameter for constructing HSS
matrix in STRUMPACK.

In this work, we mainly use two STRUMPACK driver routines: the \emph{parallel HSS construction} and
\emph{parallel HSS matrix multiplication} routines.
We use these two routines to replace the general matrix-matrix multiplication routine
\texttt{PDGEMM} in hope of achieving good speedups for large matrices.
We test the efficiency and scalability of STRUMPACK by using an HSS matrix firstly
appeared in the test routine of STRUMPACK~\cite{Strumpack}.

\textbf{Example 1}. We use two $n\times n$ Toeplitz matrices, which have been used
in~\cite{Strumpack}. In this example we assume $n=20,000$. The first one is defined as
$a_{i,i}=n^2$ and $a_{i,j}=i-j$ for $i\ne j$, which is diagonally dominant and
yields very low HSS rank (about 2).
The second one is defined as $a_{i,i}=\frac{\pi^2}{6d^2}$ and $a_{i,j}=\frac{(-1)^{i-j}}{(i-j)^2 d^2}$,
which is a kinetic energy matrix from quantum chemistry~\cite{Strumpack}, $d=0.1$ is a discretization parameter.
This matrix has slightly larger HSS rank (about $160$).

\begin{table}[ptbh]
\caption{The comparisons of HSS matrix multiplication with \texttt{PDGEMM}}
\label{tab:Ex0-time}
\begin{center}%
\begin{tabular}
[c]{|c|c|cccccc|}\hline
\multicolumn{2}{|c|}{}          & $4$    & $16$   & $64$  & $121$ & $256$ & $676$ \\ \hline \hline
\multicolumn{2}{|c|}{PDGEMM}    & 185.33 & 53.58  & 13.42 & 8.02  & 3.79  & 1.89     \\ \hline
 \multirow{3}{*}{Mat1} & Const  & 6.18   & 2.39   & 1.21  & 1.24  & 1.29  & 2.89     \\
                      & Multi   & 16.64  & 11.65  & 3.72  & 2.46  & 1.78  & 1.75     \\   \cline{2-8}
                      & Speedup & 8.12   & 3.82   & 2.72  & 2.17  & 1.23  & 0.41     \\ \hline
 \multirow{3}{*}{Mat2} & Const  & 4.57   & 1.80   & 0.96  & 1.20  & 1.23  & 2.68  \\
                      & Multi   & 15.42  & 11.10  & 3.46  & 2.29  & 1.71  & 1.58  \\   \cline{2-8}
                      & Speedup & 9.27   & 4.15   & 3.04  & 2.30  & 1.29  & 0.44 \\ \hline
\end{tabular}
\end{center}
\end{table}

From the results in Table~\ref{tab:Ex0-time}, the HSS matrix multiplication implemented in
STRUMPACK can be more than 1.2x times faster than \texttt{PDGEMM} when the used processes are
around 256.
The speedups are even better when using fewer processes.
But, the HSS construction and matrix multiplication algorithms are not as scalable as PDGEMM,
and they become slower than PDGEMM when using more processes, for example more than 676.
Therefore, it suggests to use no more than 256 processes when combining STRUMPACK with ScaLAPACK.
The execution time highly depends on many factors, such as parameters NB, block\_HSS, etc.
The results in Table~\ref{tab:Ex0-time} were obtained by choosing
$NB=64$ and $block\_HSS=512$.
Note that we did not try to find the optimal parameters.

\subsection{Combining DC with HSS}
\label{sec:hsseig}

In this section we show more details of combining HSS matrix techniques with ScaLAPACK.
As mentioned before, the central idea is to replace \texttt{PDGEMM} by the HSS matrix
multiplication algorithms.
The eigenvectors are updated in the ScaLAPACK routine \texttt{PDLAED1}, and
therefore we modify it and call STRUMPACK routines in it instead of \texttt{PDGEMM}.

Note that after applying permutations to $Q$ in~\eqref{eq:evect}, matrix $\widehat{Q}$
should also be permuted accordingly.
From the results in~\cite{Gu-eigenvalue,LSG-NLAA,Shengguo-SIMAX2}, we know that
$\widehat{Q}$ is a Cauchy-like matrix and off-diagonally low-rank, the numerical rank is usually
around $50$-$100$.
When combining with HSS, we would not use Gu's idea since permutation may destroy the
off-diagonally low-rank structure of $\widehat{Q}$ in~\eqref{eq:evect}.
We need to modify the ScaLAPACK routine \texttt{PDLAED2}, and only when the size of deflated matrix $\bar{D}$
in~\eqref{eq:T3-def} is large enough, HSS techniques are used, otherwise use Gu's idea.
Denote the size of $\bar{D}$ by $K$, and it depends the architecture of particular parallel
computers, and may be different for different computers.

Most work of DC is spent on the first two top-levels matrix-matrix multiplications.
The first top-level takes nearly 50\% when using 256 processes or fewer,
see the results in Table~\ref{tab:Ex1-time} and also the results in~\cite{Elpa}.
Therefore, we could expect at most 2x speedup when replacing \texttt{PDGEMM}
with parallel HSS matrix multiplications.

\begin{figure}[ptbh]
\centering
\subfigure[Matrix $\widehat{Q}$]{
\includegraphics[width=1.4in,height=1.4in]{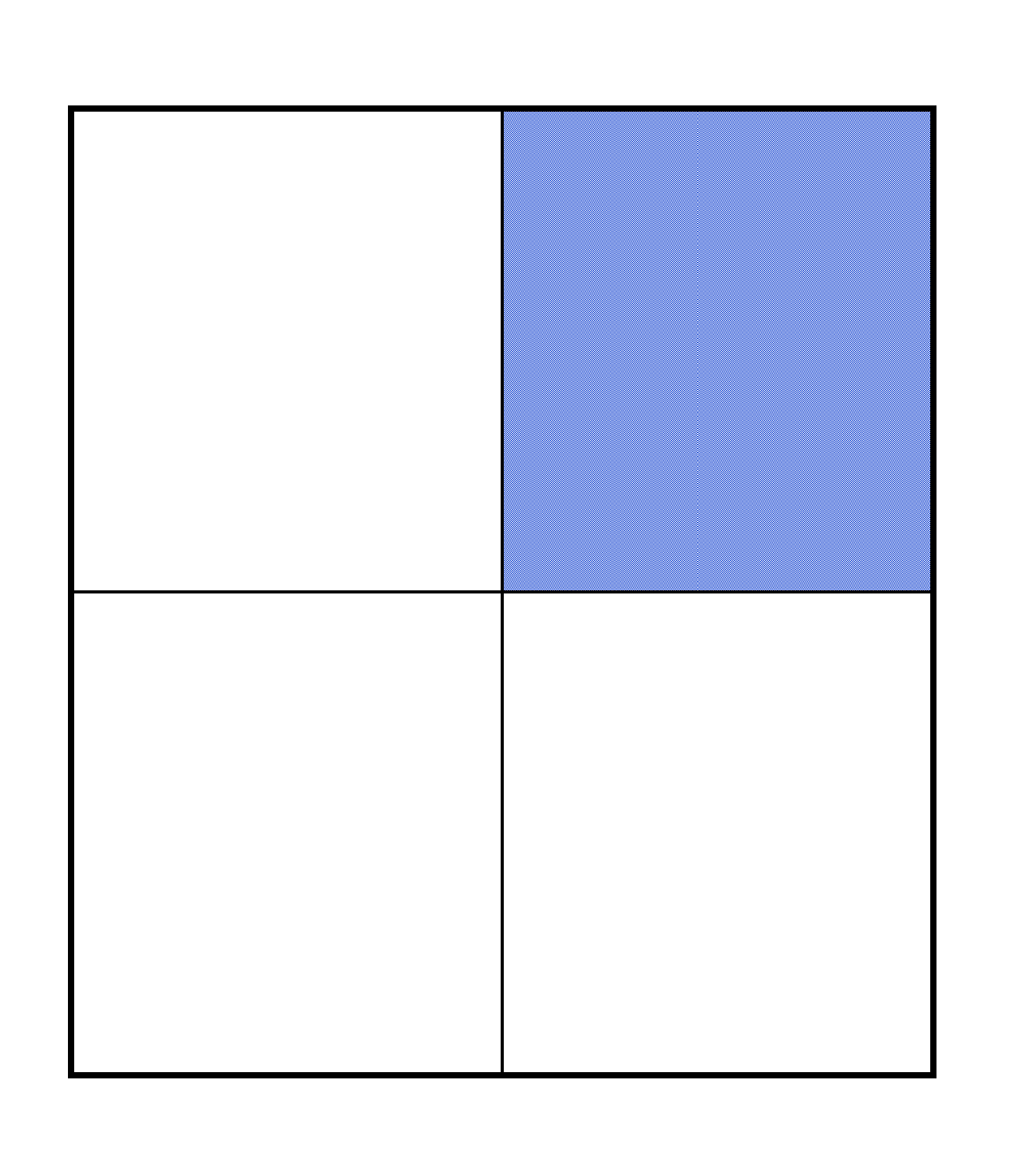}
\label{fig:matq}}
\quad
\subfigure[Singular values of $\widehat{Q}(1:m,m+1:N)$]{
\includegraphics[width=1.9in,height=1.4in]{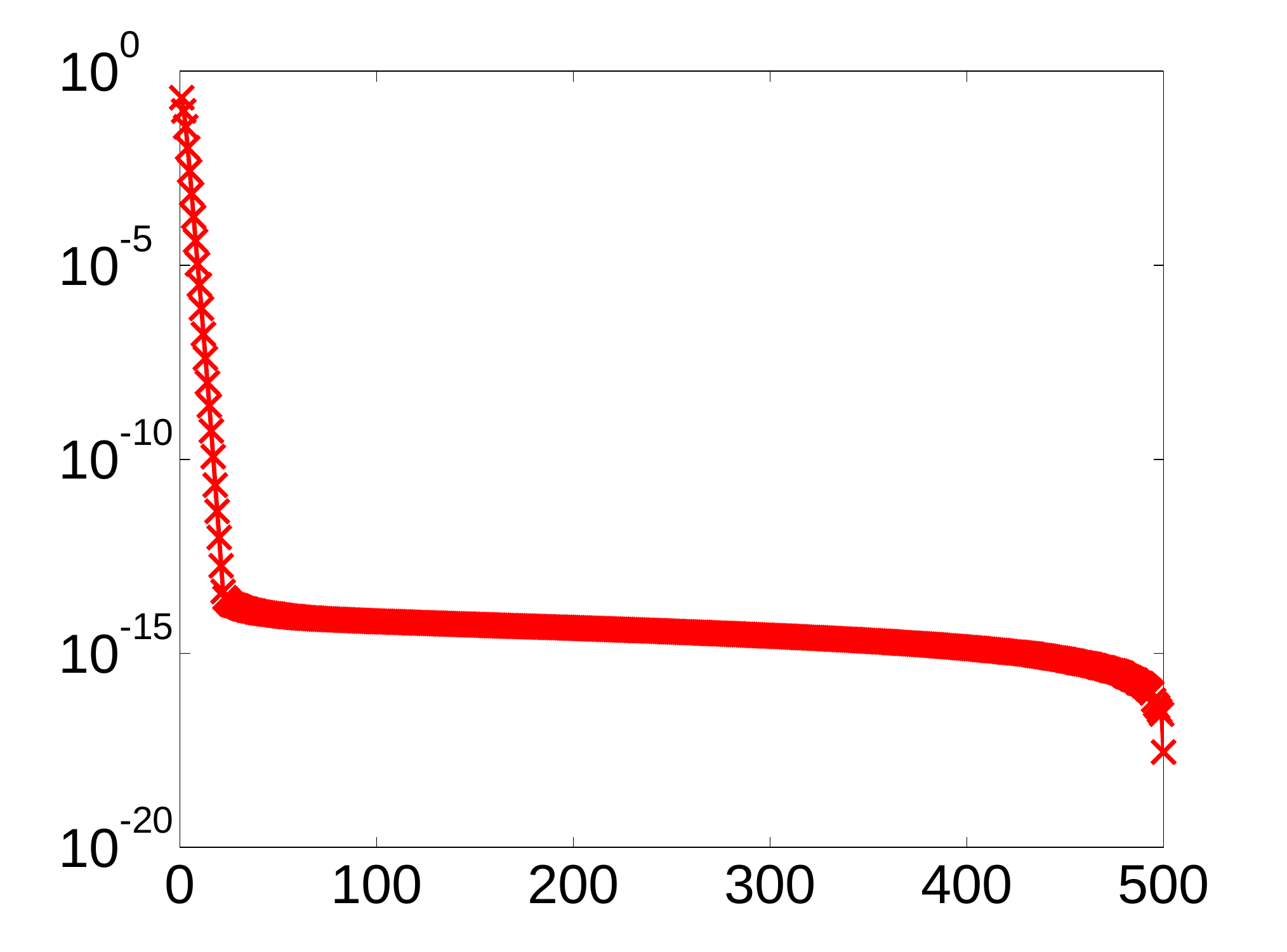}
\label{fig:svals}}%
\caption{The off-diagonally low-rank property of $\widehat{Q}$}
\end{figure}

In this subsection, we fist use an example to show the off-diagonally low-rank property of
$\widehat{Q}$ in~\eqref{eq:rankone}.
Here we assume that the dimension of $D$ is $N=1000$, and the diagonal entries of $D$
satisfy $d_i=\frac{i}{N}$, $i=1,\ldots,N$, $b_k=1$ and $u$ is a normalized
random vector in~\eqref{eq:rankone}.
Then the singular values of matrix $\widehat{Q}(1:m,m+1:N)$ are illustrated in Figure~\ref{fig:svals} with $m=500$.
From it, we get that
the singular values decay very quickly.

\textbf{Example 2.} We use a symmetric tridiagonal matrix to show the percentage of time cost
by the top level matrix-matrix multiplications.
The diagonal entries of tridiagonal matrix $A$ are all two and its off-diagonal entries are all one.
The size of this matrix is $n=20,000$.

\begin{table}[ptbh]
\caption{The percentage of time cost by the first top-level matrix-matrix multiplications}
\label{tab:Ex1-time}
\begin{center}%
\begin{tabular}
[c]{|c|cccccccc|}\hline
  & $4$ & $16$& $36$ & $64$ & $121$ & $256$ & $576$ & $1024$ \\ \hline \hline
Top One  & 59.80 & 44.13 & 30.29 & 16.92 & 9.23 & 4.77 & 1.09 & 0.53   \\
Total   & 108.04 & 77.06 & 52.78 & 30.04 & 17.70 & 10.00 & 6.20 & 6.22 \\ \hline
Percent(\%) & 55.35 & 57.2 & 57.39 & 56.32 & 52.15 & 47.70 & 17.58 & 8.52 \\ \hline
\end{tabular}
\end{center}
\end{table}

The results in Table~\ref{tab:Ex1-time} are obtained by using optimization flags
\texttt{"-O2 -C -qopenmp -mavx"}, and linked with multi-threaded Intel MKL.
From the results in it, we can see that the top-one level matrix-matrix multiplications can take
half of the total time cost by \texttt{PDSTEDC} in some case.
Since \texttt{PDGEMM} in MKL has very good scalability, the percentage of top-one level matrix multiplications
decreases as the number of processes increases.
This example also implies that we are better not to use more than 256 processes
in our numerical experiments.

\section{Numerical results}
\label{sec:num}

All the results are obtained on Tianhe-2 supercomputer~\cite{Liao-TH2, Liao-HPCG},
located in Guangzhou, China. It employs accelerator
based architectures and each compute node is equipped with two Intel
Xeon E5-2692 CPUs and three Intel Xeon Phi accelerators based on the
many-integrated-core (MIC) architectures. In our experiments we only
use CPU cores.

\textbf{Example 3.} We use some `difficult' matrices~\cite{A880} for the DC algorithm,
for which few or no eigenvalues are deflated. Examples include
the Clement-type, Hermite-type and Toeplitz-type matrices, which are defined as follows.

The Clement-type matrix~\cite{A880} is given by
\begin{equation}
  \label{eq:Clement-Tri}
  T=\text{tridiag}
  \begin{pmatrix}
    &\sqrt{n} & & \sqrt{2(n-1)} & & \sqrt{(n-1)2} & & \sqrt{n} & \\
    0 & & 0 & & \ldots & & 0 & & 0 \\
    &\sqrt{n} & & \sqrt{2(n-1)} & & \sqrt{(n-1)2} & & \sqrt{n} & \\
  \end{pmatrix},
\end{equation}
where the off-diagonal entries are $\sqrt{i(n+1-i)}, i=1,\ldots,n$.

The Hermite-type matrix is given as~\cite{A880},
\begin{equation}
  \label{eq:Hermite-Tri}
  T=\text{tridiag}
  \begin{pmatrix}
    &\sqrt{1} & & \sqrt{2} & & \sqrt{n-2} & & \sqrt{n-1} & \\
    0 & & 0 & & \ldots & & 0 & & 0 \\
    &\sqrt{1} & & \sqrt{2} & & \sqrt{n-2} & & \sqrt{n-1} & \\
  \end{pmatrix}.
\end{equation}

The Toeplitz-type matrix is defined as~\cite{A880},
\begin{equation}
  \label{eq:Laguerre-Tri}
  T=\text{tridiag}
  \begin{pmatrix}
    &1 & & 1 & & 1 & & 1 & \\
    2 & & 2 & & \ldots & & 2 & & 2 \\
    &1 & & 2 & & 1 & & 1 & \\
  \end{pmatrix}.
\end{equation}

\begin{figure}[ptbh]
\centering
\subfigure[The strong scaling of PHDC]{
\includegraphics[width=2.4in,height=2.2in]{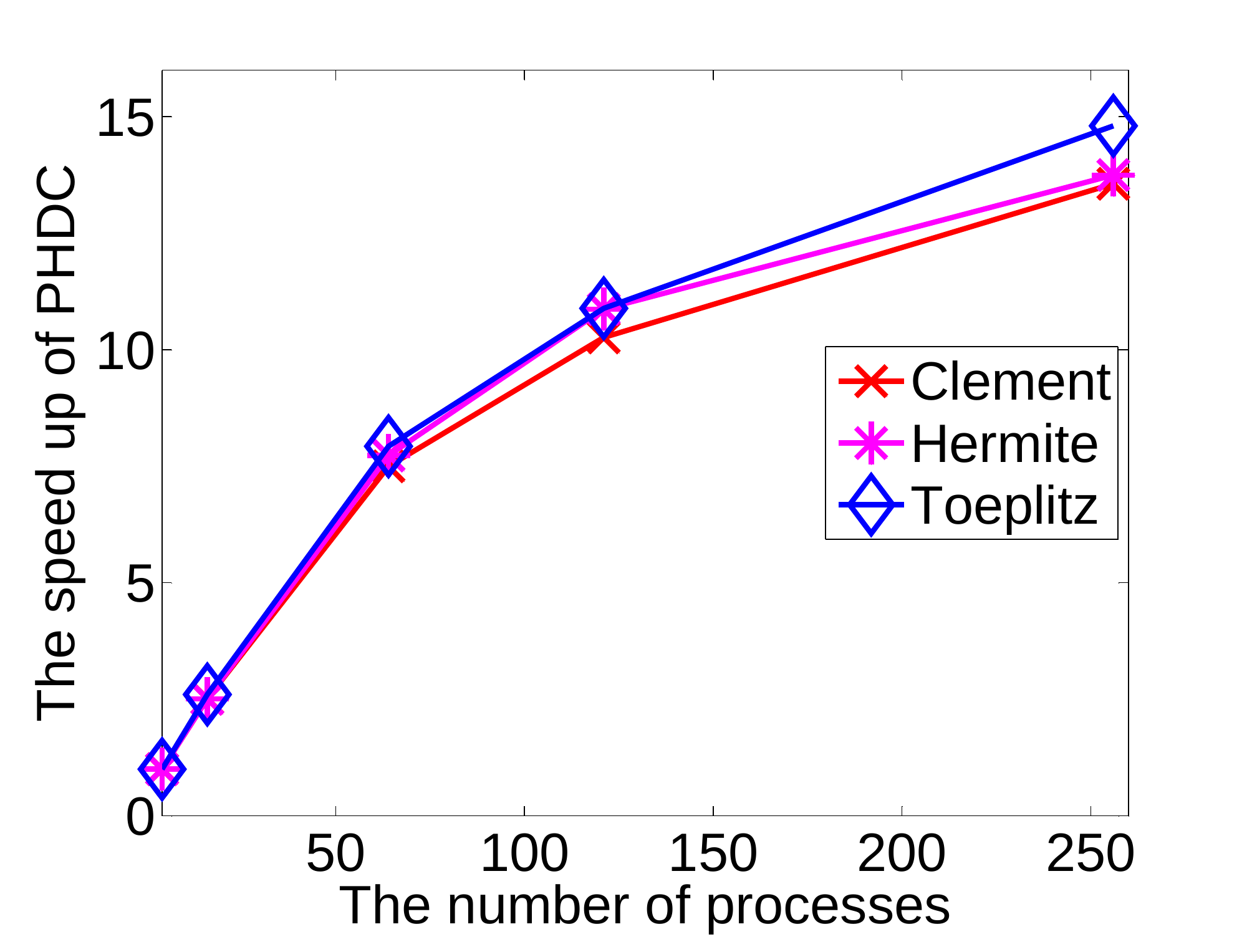}
\label{fig:strong}}
\subfigure[The speedups of PHDC over ScaLAPACK]{
\includegraphics[width=2.4in,height=2.3in]{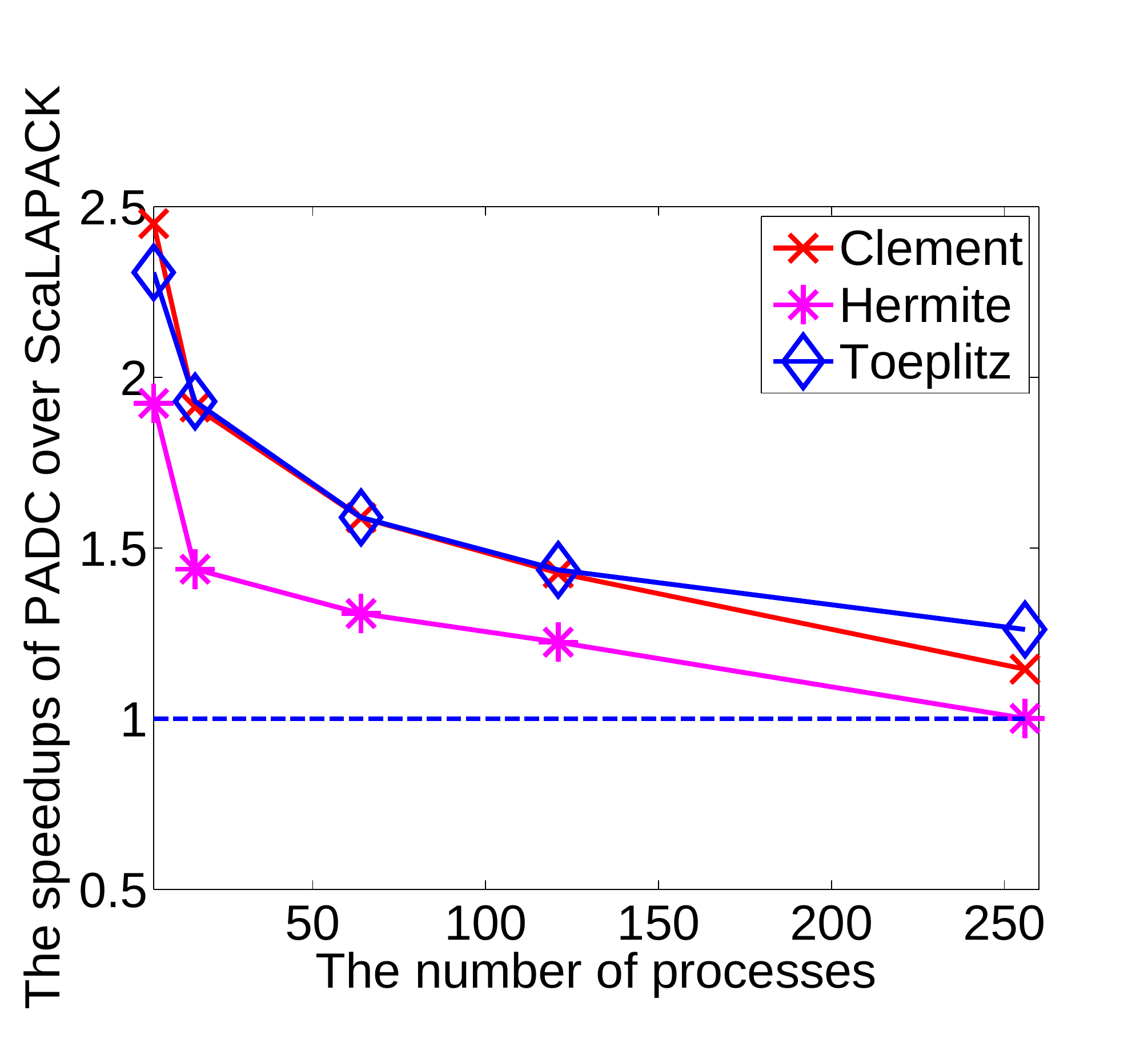}
\label{fig:padc-scal}}
\caption{The results for the difficult matrices}
\label{fig:padc-ex3}
\end{figure}

For the results of strong scaling, we let the dimension be $n$=$30,000$, and use
HSS techniques only when the size of secular equation is larger than $K=20,000$.
The results for strong scaling are shown in Figure~\ref{fig:strong}.
The speedups of PHDC over ScaLAPACK are reported in Table~\ref{tab:Ex3-strong},
and shown in Figure~\ref{fig:padc-scal}.
We can see that PHDC is about $1.4$ times faster than PDSTEDC in MKL when
using 120 processes or fewer.

\begin{table}[ptbh]
\caption{The strong scaling of PHDC compared with Intel MKL}
\label{tab:Ex3-strong}
\begin{center}%
\begin{tabular}
[c]{|c|ccccc|}\hline
\multirow{2}{*}{Matrix}  & \multicolumn{5}{c|}{Number of Processes} \\ \cline{2-6}
  & $4$ & $16$& $64$ & $121$ & $256$ \\ \hline \hline
Clement  & 2.45 & 1.91 & 1.58 & 1.42 & 1.14    \\
Hermite  & 1.92& 1.43 & 1.30 & 1.22& 1.00   \\
Toeplitz & 2.30 & 1.92 & 1.58 & 1.43 & 1.26  \\ \hline
\end{tabular}
\end{center}
\end{table}

The orthogonality of the computed eigenvectors by PHDC are
in the same order as those by ScaLAPACK,
which are shown in Table~\ref{tab:Ex3-orth}.
The orthogonality of matrix $Q$ is defined as $\|I-QQ^T\|_{\max}$,
where $\| \bullet \|_{\max}$ is the maximum absolute value of entries of $(\bullet)$.

\begin{table}[ptbh]
\caption{The orthogonality of the computed eigenvectors by PHDC}
\label{tab:Ex3-orth}
\begin{center}%
\begin{tabular}
[c]{|c|ccccc|}\hline
\multirow{2}{*}{Matrix}  & \multicolumn{5}{c|}{Number of Processes} \\ \cline{2-6}
  & $4$ & $16$& $64$ & $121$ & $256$ \\ \hline \hline
Clement  & $1.6e$-$13$ & $1.6e$-$13$ & $1.6e$-$13$ & $1.6e$-$13$ & $1.6e$-$13$   \\
Hermite  & $3.6e$-$13$ & $3.6e$-$13$ & $3.6e$-$13$ & $3.6e$-$13$ & $3.6e$-$13$   \\
Toeplitz & $2.9e$-$13$ & $2.9e$-$13$ & $2.9e$-$13$ & $2.9e$-$13$ & $2.9e$-$13$  \\ \hline
\end{tabular}
\end{center}
\end{table}

\textbf{Example 4.} In~\cite{Tygert-SHT2}, Tygert shows that the spherical harmonic transform (SHT) can be
accelerated using the tridiagonal DC algorithm and one symmetric tridiagonal matrix is defined as
follows,
\begin{equation}
A_{jk}=\begin{cases} c_{m+2j-2}, & k=j-1 \\ d_{m+2j}, & k=j \\ c_{m+2j}, & k=j+1 \\ 0, & otherwise, \end{cases}
\end{equation}
for $j, k=0,1,\ldots,n-1$, where
$$c_l=\sqrt{\frac{(l-m+1)(l-m+2)(l+m+1)(l+m+2)}{(2l+1)(2l+3)^2(2l+5)}}, \quad d_l=\frac{2l(l+1)-2m^2-1}{(2l-1)(2l+3)},$$
for $l=m, m+1, m+2,\ldots.$
Assume that the dimension of this matrix is $n=30,000$ and $m=n$.
The execution times of using PHDC and PDSTEDC are reported in Table~\ref{tab:Ex4-time}.

\begin{table}[ptbh]
\caption{The execution time of PHDC and PDSTEDC for SHT}
\label{tab:Ex4-time}
\begin{center}%
\begin{tabular}
[c]{|c|ccccc|}\hline
\multirow{2}{*}{Method}  & \multicolumn{5}{c|}{Number of Processes} \\ \cline{2-6}
  & $4$ & $16$& $64$ & $121$ & $256$ \\ \hline \hline
PDSTEDC  & 475.00 & 139.09 & 39.66 & 25.82 & 16.87   \\
PHDC     & 224.97 & 88.58 & 28.53 & 20.12 & 14.71    \\ \hline
Speedup  & 2.11 & 1.57 & 1.39 & 1.28 & 1.15    \\ \hline
\end{tabular}
\end{center}
\end{table}

\subsection{Comparison with other implementations}
\label{sec:elpa}

Different from ScaLAPACK, the ELPA routines~\cite{Elpa,elpa-library} do not rely on BLACS, all
communication between different processors is handled by direct calls to a
MPI library, where ELPA stands for \emph{Eigenvalue soLver for Petascale Applications}.
For its communications, ELPA relies on two separate sets of MPI communicators,
row communicators and column communicators, respectively (connecting either
the processors that handle the same rows or the same columns).
For the tridiagonal eigensolver, ELPA implements its own matrix-matrix multiplications,
does not use PBLAS routine \texttt{PDGEMM}.
It is known that ELPA has better scalability and is faster than MKL~\cite{Elpa,elpa-library}.

\begin{figure}[ptbh]
\centering
\includegraphics[width=3.0in,height=2.4in]{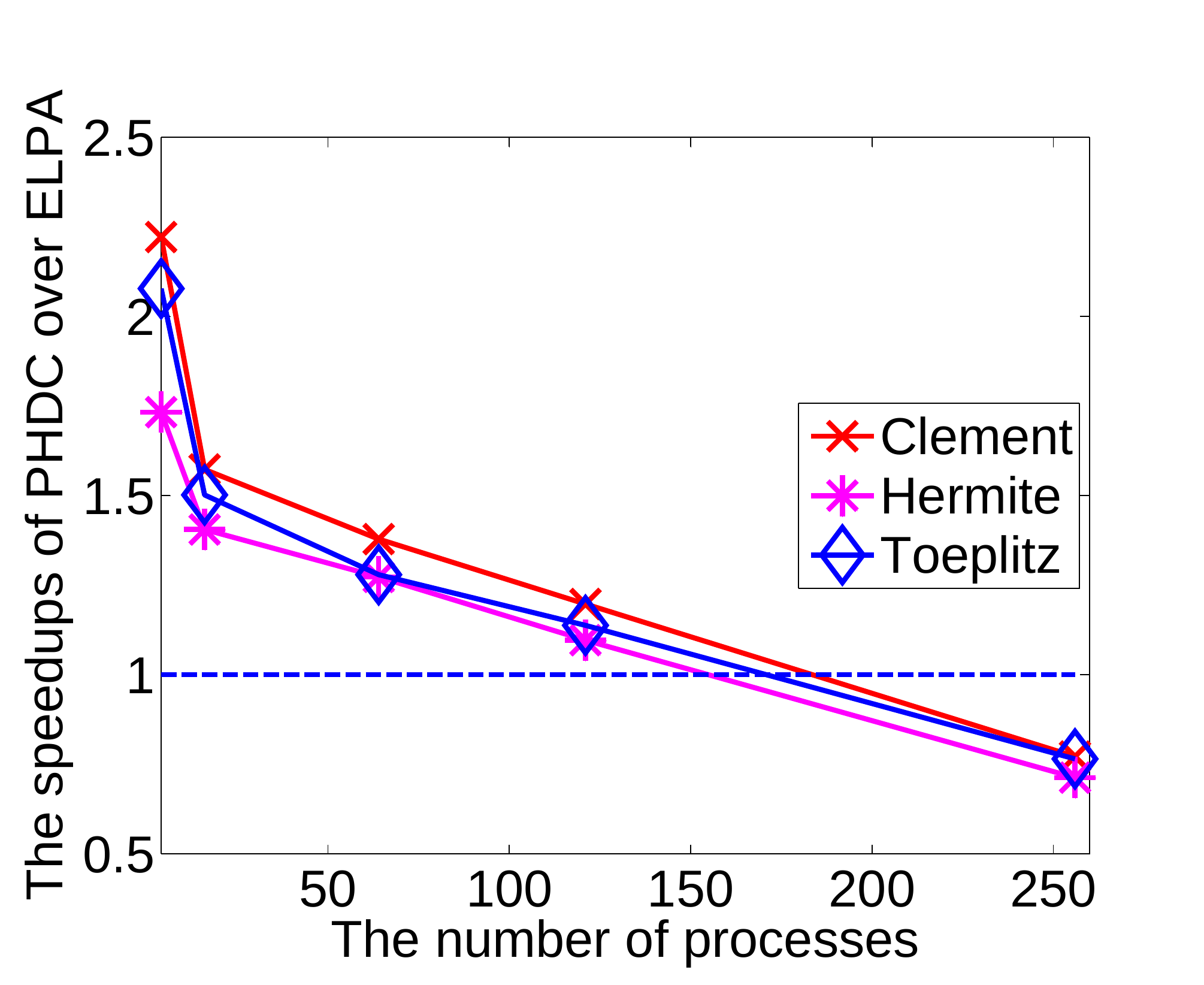}
\caption{The speedups of PHDC over ELPA for the difficult matrices}
\label{fig:spd-elpa}
\end{figure}

\textbf{Example 5.} We use the same matrices as in Example 3 to test ELPA, and compare it with
the newly proposed algorithm PHDC.
The running times of ELPA are shown in Table~\ref{tab:time-elpa}.
Figure~\ref{fig:spd-elpa} shows the speedups of PHDC over ELPA.
PHDC is faster than ELPA since it requires fewer floating point operations.
However, its scalability is worse than ELPA, and PHDC becomes slower than
ELPA when using more than 200 processes.

\begin{table}[ptbh]
\caption{The execution time of ELPA for different matrices}
\label{tab:time-elpa}
\begin{center}%
\begin{tabular}
[c]{|c|ccccc|}\hline
\multirow{2}{*}{Matrix}  & \multicolumn{5}{c|}{Number of Processes} \\ \cline{2-6}
  & $4$ & $16$& $64$ & $121$ & $256$ \\ \hline \hline
Clement  & 487.63 & 137.43 & 40.36 & 25.61 & 12.49   \\
Hermite  & 363.94 & 117.61 & 34.58 & 21.18 & 10.89  \\
Toeplitz & 509.32 & 141.57 & 39.55 & 25.62 & 12.67  \\ \hline
\end{tabular}
\end{center}
\end{table}

\section{Conclusions}
\label{sec:conclusion}

By combining ScaLAPACK with STRUMPACK, we propose a hybrid tridiagonal DC algorithm for
the symmetric eigenvalue problems, which can be faster than the classical DC algorithm implemented in ScaLAPACK when
using about 200 processes.
The central idea is to relpace \texttt{PDGEMM} by the HSS matrix multiplication algorithms,
since HSS matrix algorithms require fewer flops than \texttt{PDGEMM}.
Numerical results show that the scalability of HSS matrix algorithms is not as good as \texttt{PDGEMM}.
The proposed PHDC algorithm in this work becomes slower than the classical DC algorithm
when using more processes.

\section*{Acknowledgement}
The authors would like to acknowledge many helpful discussions with Xiangke Liao, Sherry Li and Shuliang Lin.
This work is partially supported by National Natural Science Foundation of China (Nos. 11401580, 91530324,
91430218 and 61402495).


\begin{thebibliography}{10}

\bibitem{anderson1999lapack}
E.~Anderson, Z.~Bai, C.~Bischof, S.~Blackford, J.~Demmel, J.~Dongarra,
  J.~Du~Croz, A.~Greenbaum, S.~Hammarling, A.~McKenney, and D.~Sorensen.
\newblock {\em {LAPACK} Users' Guide}.
\newblock Society for Industrial and Applied Mathematics, Philadelphia, PA,
  third edition, 1999.

\bibitem{Elpa}
T.~Auckenthaler, V.~Blum, H.~J. Bungartz, T.~Huckle, R.~Johanni, L.~Kr\"{a}mer,
  B.~Lang, H.~Lederer, and P.~R. Willems.
\newblock Parallel solution of partial symmetric eigenvalue problems from
  electronic structure calculations.
\newblock {\em Parallel Computing}, 37(12):783--794, 2011.

\bibitem{Hlib}
S.~B\"{o}rm and L.~Grasedyck.
\newblock {H-Lib}-- a library for $\mathcal{H}$- and $\mathcal{H}^2$-matrices,
  1999.

\bibitem{BNS-Rankone}
J.~R. Bunch, C.~P. Nielsen, and D.~C. Sorensen.
\newblock Rank one modification of the symmetric eigenproblem.
\newblock {\em Numer. Math.}, 31:31--48, 1978.

\bibitem{Chandrasekaran03}
S.~Chandrasekaran, P.~Dewilde, M.~Gu, T.~Pals, X.~Sun, A.~J. van~der Veen, and
  D.~White.
\newblock Fast stable solvers for sequentially semi-separable linear systems of
  equations and least squares problems.
\newblock Technical report, University of California, Berkeley, CA, 2003.

\bibitem{ChandrasekaranGu05}
S.~Chandrasekaran, P.~Dewilde, M.~Gu, T.~Pals, X.~Sun, A.~J. van~der Veen, and
  D.~White.
\newblock Some fast algorithms for sequentially semiseparable representation.
\newblock {\em SIAM J. Matrix Aanal. Appl.}, 27:341--364, 2005.

\bibitem{CG-SSS-report}
S.~Chandrasekaran, P.~Dewilde, M.~Gu, T.~Pals, and A.~J. van~der Veen.
\newblock Fast stable solvers for sequentially semi-separable linear systems of
  equations.
\newblock Technical report, UC, Berkeley, CA, 2003.

\bibitem{ChandrasekaranGu04}
S.~Chandrasekaran, M.~Gu, and T.~Pals.
\newblock Fast and stable algorithms for hierarchically semi-separable
  representations.
\newblock Technical report, University of California, Berkeley, CA, 2004.

\bibitem{Hss-ulv}
S.~Chandrasekaran, M.~Gu, and T.~Pals.
\newblock A fast {ULV} decomposition solver for hierarchical semiseparable
  representations.
\newblock {\em SIAM J. Matrix Anal. Appl.}, 28:603--622, 2006.

\bibitem{SSS-QR}
S.~Chandrasekaran, M.~Gu, J.~Xia, and J.~Zhu.
\newblock A fast {QR} algorithm for companion matrices.
\newblock In I.~A. Ball and .et al., editors, {\em Recent Advances in Matrix
  and Operator Theory}, pages 111--143, Birkh\''{a}user, Basel, 2008.

\bibitem{Cheng-Random}
H.~Cheng, Z.~Gimbutas, P.~Martinsson, and V.~Rokhlin.
\newblock On the compression of low rank matrices.
\newblock {\em SIAM J. Sci. Comput.}, 26(4):1389--1404, 2005.

\bibitem{Scalapack}
J.~Choi, J.~Demmel, I.~Dhillon, J.~Dongarra, S.~Ostrouchov, A.~Petitet,
  K.~Stanley, D.~Walker, and R.C. Whaley.
\newblock Scalapack: A portable linear algebra library for distributed memory
  computers-design issues and performance.
\newblock {\em Computer Physics Communications}, 97:1--15, 1996.

\bibitem{Cuppen81}
J.~J.~M. Cuppen.
\newblock A divide and conquer method for the symmetric tridiagonal
  eigenproblem.
\newblock {\em Numer. Math.}, 36:177--195, 1981.

\bibitem{Eidelman-Gohberg1999}
Y.~Eidelman and I.~Gohberg.
\newblock On a new class of structured matrices.
\newblock {\em Integral Equations and Operator Theory}, 34:293--324, 1999.

\bibitem{Gates-Arbenz}
K.~Gates and P.~Arbenz.
\newblock Parallel divide and conquer algorithms for the symmetric tridiagonal
  eigenproblem.
\newblock Technical report, Institute for Scientific Comouting, ETH Zurich,
  Zurich, Switzerland, 1994.

\bibitem{Strumpack-sparse}
P.~Ghysels, X.~Li, and F.~Rouet~S. Williams.
\newblock An efficient multi-core implementation of a novel {HSS}-structured
  multifrontal solver using randomized sampling, 2014.
\newblock Submitted to SIAM J. Sci. Comput.

\bibitem{Gu-thesis}
M.~Gu.
\newblock {\em Studies in Numerical Linear Algebra}.
\newblock PhD thesis, Yale University, New Haven, CT, 1993.

\bibitem{Gu-eigenvalue}
M.~Gu and S.~C. Eisenstat.
\newblock A divide-and-conquer algorithm for the symmetric tridiagonal
  eigenproblem.
\newblock {\em SIAM J. Matrix Anal. Appl.}, 16:172--191, 1995.

\bibitem{Gu-DSSS}
M.~Gu, X.~S. Li, and P.~S. Vassilevski.
\newblock Direction-preserving and schur-monotonic semiseparable approximations
  of symmetric positive definite matrices.
\newblock {\em SIAM J. Matrix Anal. Appl.}, 31:2650--2664, 2010.

\bibitem{Hackbusch1999}
W.~Hackbusch.
\newblock A sparse matrix arithmetic based on $\mathcal{H}$-matrices. {Part I}:
  Introduction to $\mathcal{H}$-matrices.
\newblock {\em Computing}, 62:89--108, 1999.

\bibitem{Hackbusch-Borm2002}
W.~Hackbusch and S.~B\"{o}rm.
\newblock Data-sparse approximation by adaptive $\mathcal{H}^2$-matrices.
\newblock {\em Computing}, 69:1--35, 2002.

\bibitem{Hackbusch-Sauter2000}
W.~Hackbusch, B.~Khoromskij, and S.~Sauter.
\newblock On $\mathcal{H}^2$-matrices.
\newblock In Zenger~C Bungartz~H, Hoppe~RHW, editor, {\em Lecture on Applied
  Mathematics}, pages 9--29, Berlin, 2000. Springer.

\bibitem{Hackbusch2000}
W.~Hackbusch and B.N. Khoromskij.
\newblock A sparse matrix arithmetic based on $\mathcal{H}$-matrices. {Part
  II}: Application to multi-dimensional problems.
\newblock {\em Computing}, 64:21--47, 2000.

\bibitem{Martinsson-Rev10}
N.~Halko, P.~G. Martinsson, and J.~A. Tropp.
\newblock Finding structure with randomness probabilistic algorithms for
  constructing approximate matrix decompositions.
\newblock {\em SIAM Review}, 53:217--288, 2011.

\bibitem{Ipsen-TDC}
I.~Ipsen and E.~Jessup.
\newblock Solving the symmetric tridiagonal eigenvalue problem on the
  hypercube.
\newblock {\em SIAM J. Sci. Statist. Comput.}, 11:203--229, 1990.

\bibitem{Shengguo-SIMAX2}
S.~Li, M.~Gu, L.~Cheng, X.~Chi, and M.~Sun.
\newblock An accelerated divide-and-conquer algorithm for the bidiagonal {SVD}
  problem.
\newblock {\em SIAM J. Matrix Anal. Appl.}, 35(3):1038--1057, 2014.

\bibitem{LSG-NLAA}
S.~Li, X.~Liao, J.~Liu, and H.~Jiang.
\newblock New fast divide-and-conquer algorithm for the symmetric tridiagonal
  eigenvalue problem.
\newblock {\em Numer. Linear Algebra Appl.}, 23:656--673, 2016.

\bibitem{Liao-camwa}
X.~Liao, S.~Li, L.~Cheng, and M.~Gu.
\newblock {An improved divide-and-conquer algorithm for the banded matrices
  with narrow bandwidths}.
\newblock {\em Comput. Math. Appl.}, 71:1933--1943, 2016.

\bibitem{Liao-TH2}
X.~Liao, L.~Xiao, C.~Yang, and Y.~Lu.
\newblock Milkyway-2 supercomputer: System and application.
\newblock {\em Frontiers of Computer Science}, 8(3):345--356, 2014.

\bibitem{Martinsson-PNAS07}
E.~Liberty, F.~Woolfe, P.~G. Martinsson, V.~Rokhlin, and M.~Tygert.
\newblock Randomized algorithms for the low-rank approximation of matrices.
\newblock {\em PNAS}, 104(51):20167--20172, 2007.

\bibitem{Liao-HPCG}
Y.~Liu, C.~Yang, F.~Liu, X.~Zhang, Y.~Lu, Y.~Du, C.~Yang, M.~Xie, and X.~Liao.
\newblock {623 Tflop/s HPCG} run on tianhe-2: Leveraging millions of hybrid
  cores.
\newblock {\em International Journal of High Performace Computing
  Applications}, 30(1):39--54, 2016.

\bibitem{Lyons-thesis}
W.~Lyons.
\newblock {\em Fast algorithms with applications to {PDEs}}.
\newblock PhD thesis, University of California, Santa Barbara, 2005.

\bibitem{elpa-library}
A.~Marek, V.~Blum, R.~Johanni, V.~Havu, B.~Lang, T.~Auckenthaler, A.~Heinecke,
  H.~Bungartz, and H.~Lederer.
\newblock {The ELPA library:} scalable parallel eigenvalue solutions for
  electronic structure theory and computational science.
\newblock {\em J. Phys.: Condens. Matter}, 26:1--15, 2014.

\bibitem{A880}
O.~A. Marques, C.~Voemel, J.~W. Demmel, and B.~N. Parlett.
\newblock Algorithm 880: A testing infrastructure for symmetric tridiagonal
  eigensolvers.
\newblock {\em ACM Trans. Math. Softw.}, 35:8:1--13, 2008.

\bibitem{Martinsson-randhss10}
P.~G. Martinsson.
\newblock A fast randomized algorithm for computing a hierarchically
  semiseparable representation of a matrix.
\newblock {\em SIAM J. Matrix Anal. Appl.}, 32:1251--1274, 2011.

\bibitem{Strumpack}
F.~Rouet, X.~Li, P.~Ghysels, and A.~Napov.
\newblock A distributed-memory package for dense hierarchically semi-separable
  matrix computations using randomization.
\newblock {\em ACM Transactions on Mathematical Software}, 42(4):27:1--27:35,
  2016.

\bibitem{Sorensen-sina91}
D.~C. Sorensen and P.~T.~P. Tang.
\newblock On the orthogonality of eigenvectors computed by divide-and-conquer
  techniques.
\newblock {\em SIAM J. Numer. Anal.}, 28:1752--1775, 1991.

\bibitem{Tisseur-DC}
F.~Tisseur and J.~Dongarra.
\newblock A parallel divide and conquer algorithm for the symmetric eigenvalue
  problem on distributed memory architectures.
\newblock {\em SIAM J. Sci. Comput.}, 20(6):2223--2236, 1999.

\bibitem{Tygert-SHT2}
M.~Tygert.
\newblock Fast algorithms for spherical harmonic expansions, ii.
\newblock {\em Journal of Computational Physics}, 227:4260--4279, 2008.

\bibitem{Vandebril-book1}
R.~Vandebril, M.~Van~Barel, and N.~Mastronardi.
\newblock {\em Matrix Computations and Semiseparable Matrices, {Volume I}:
  Linear Systems}.
\newblock Johns Hopkins University Press, 2008.

\bibitem{Xia-Fast09}
J.~Xia, S.~Chandrasekaran, M.~Gu, and X.S. Li.
\newblock Fast algorithm for hierarchically semiseparable matrices.
\newblock {\em Numer. Linear Algebra Appl.}, 17:953--976, 2010.

\bibitem{Xia-HSS-Chol}
J.~Xia and M.~Gu.
\newblock Robust approximate {Choleksy} factorization of rank-structured
  symmetric positive definite matrices.
\newblock {\em SIAM J. Matrix Anal. Appl.}, 31:2899--2920, 2010.

\bibitem{Xia-random}
J.~Xia, Y.~Xi, and M.~Gu.
\newblock A superfast structured solver for {Toeplitz} linear systems via
  randomized sampling.
\newblock {\em SIAM J. Matrix Anal. Appl.}, 33:837--858, 2012.

\end{thebibliography}
\end{document}